\documentstyle[12pt]{article}
 \makeatletter
 \def\leqq{\mathrel{\mathpalette\gl@align<}}
 \def\geqq{\mathrel{\mathpalette\gl@align>}}
 \def\gl@align#1#2{\lower.6ex\vbox{\baselineskip\z@skip\lineskip\z@
     \ialign{$\m@th#1\hfil##\hfil$\crcr#2\crcr=\crcr}}}
 \makeatother

 \makeatletter
 \def\sileqq{\mathrel{\mathpalette\gs@align<}}
 \def\sigeqq{\mathrel{\mathpalette\gs@align>}}
 \def\gs@align#1#2{\lower.6ex\vbox{\baselineskip\z@skip\lineskip\z@
     \ialign{$\m@th#1\hfil##\hfil$\crcr#2\crcr\sim\crcr}}}
 \makeatother
\begin{document}
\hbadness=10000
\hbadness=10000
\begin{titlepage}
\nopagebreak
\begin{flushright}
\end{flushright}
\vspace{1.5cm}
\begin{center}

{\large \bf Generalized Matrix Mechanics}

\vspace{1cm}
 
{Yoshiharu Kawamura}\footnote{E-mail:
haru@azusa.shinshu-u.ac.jp} 

\vspace{0.7cm}
Department of Physics, Shinshu University,
Matsumoto 390-8621, Japan\\
\end{center}
\vspace{0.7cm}

\nopagebreak

\begin{abstract}
We propose a generalization of Heisenberg's matrix mechanics
based on many-index objects.
It is shown that there exists a solution describing a harmonic oscillator
and that the many-index objects lead to a generalization of spin algebra.
\end{abstract}
\vfill
\end{titlepage}
\pagestyle{plain}
\newpage
\def\thefootnote{\fnsymbol{footnote}}

\section{Introduction}

Until the end of 19th century, it was generally believed
that any experimental results could be explained with
classical mechanics (CM).
The phenomenon of black body radiation destroyed this belief,
and the concept of energy quanta was introduced by Planck
in 1900 to overcome the difficulty presented.
Since that time, quantum mechanics (QM) has been applied to very broad
areas of physics with indisputable success.
Considering its success, it is natural to ask the following questions:
\begin{enumerate}
\item Why does QM describe the microscopic world so successfully?
\item Does QM hold without limit?
\item If there are limitations, how is QM modified beyond them?
\end{enumerate}
Unfortunately, we presently have no definite answers to these questions, although
there are some conjectures. 
We expect that a generalization of CM and/or QM
will provide information that can help to answer the above questions. From
this point of view, it is meaningful to construct a new, generalized 
mechanics based on CM and/or QM.

Nambu proposed a generalization of Hamiltonian dynamics through the extension
of phase space based on the Liouville theorem and gave a suggestion
for its quantization.\cite{Nambu}
The structure of this mechanics has been studied in the framework of constrained
systems \cite{con} and in geometric and algebraic formulations.\cite{geo}
There are several works in which the quantization of Nambu mechanics
is investigated.\cite{geo,q1,q2,q3,q4,q5}
This approach is quite interesting, but it is not the unique way to explore
new mechanics.
There is also the possibility of examining the generalization of QM directly, 
and here we consider this possibility.

In this paper, we propose a generalization of Heisenberg's matrix mechanics
based on many-index objects (which we refer to as the M-matrix).\footnote{
Recently, Awata, Li, Minic and Yoneya introduced many-index objects
to quantize Nambu mechanics.\cite{q3}
We find that our definition of the triple product among cubic matrices 
is different from theirs, because we require a generalization of the Ritz rule
in the phase factor, but not necessarily the associativity of the products.}
It is shown that there exists a solution describing a harmonic oscillator
and that the many-index objects lead to a generalization of spin algebra.
A conjecture concerning operator formalism is also given.

This paper is organized as follows. 
In the next section, we review Heisenberg's matrix mechanics
and explore its generalization.
We formulate (cubic) matrix mechanics based on three-index objects in $\S$3. 
Section 4 is devoted to conclusions and discussion.

\section{Matrix mechanics and generalization}
 \subsection{Heisenberg's matrix mechanics}

Here we review Heisenberg's matrix mechanics.
For a closed physical system, physical quantities are represented by hermitian square matrices that
can be written as
\begin{eqnarray}
F_{mn}(t) = F_{mn} e^{i\Omega_{mn}t} = F_{mn} e^{{i \over \hbar}(E_m -E_n)t} ,
\label{Fmn}
\end{eqnarray}
where the phase factor implies that a change in energy $E_{m} - E_n$ appears as 
radiation with angular frequency $\Omega_{mn}$, and the hermiticity of $F_{mn}(t)$
is expressed by $F_{nm}^{*}(t) = F_{mn}(t)$.
By the usual definition of the product of two square matrices $A_{mn}(t) = A_{mn} e^{i\Omega_{mn}t}$ 
and $B_{mn}(t) = B_{mn} e^{i\Omega_{mn}t}$,
\begin{eqnarray}
(AB)_{mn}(t) \equiv \sum_k A_{mk}(t) B_{kn}(t) = \sum_k A_{mk} B_{kn} e^{i\Omega_{mn}t} ,
\label{ABmn}
\end{eqnarray}
it is seen that the product $(AB)_{mn}(t)$ has the same form as (\ref{Fmn}), with
the Ritz rule $\Omega_{mn} = \Omega_{mk} + \Omega_{kn}$.
The time development of $F_{mn}(t)$ is expressed by the Heisenberg equation
\begin{eqnarray}
{d \over dt}F_{mn}(t) &=& i \Omega_{mn} F_{mn} = {i \over \hbar}(E_m - E_n)F_{mn}(t) \nonumber \\
&=& {1 \over i\hbar} ((F(t)H)_{mn} - (HF(t))_{mn}) \equiv {1 \over i\hbar} [F(t), H]_{mn} ,
\label{H-eq}
\end{eqnarray} 
where the Hamiltonian $H$ is a diagonal matrix written $H_{mn} \equiv E_m \delta_{mn}$.

Here we give a simple example of a harmonic oscillator whose variables are two hermitian matrices,
$\xi_{mn}(t) = \xi_{mn} e^{i\Omega_{mn}t}$ and 
$\eta_{mn}(t) = \eta_{mn} e^{i\Omega_{mn}t}$.
The coefficients $\xi_{mn}$ and $\eta_{mn}$ are given by
\begin{eqnarray}
\xi_{mn} = \sqrt{{\hbar \over 2m\Omega}} (\sigma^1)_{mn}  ~~\mbox{and}~~
\eta_{mn} = \sqrt{{m \Omega \hbar \over 2}} (\sigma^2)_{mn} ,
\label{xieta}
\end{eqnarray}
respectively.
Here the quantity $m$ in the square root represents a mass, the $(\sigma^a)_{mn}$ are Pauli matrices,
and $\Omega = \Omega_{21}(>0)$.
The variables $\xi_{mn}(t)$ and $\eta_{mn}(t)$ satisfy the following anticommutation relations:
\begin{eqnarray}
&~& \{\xi(t), \xi(t)\}_{mn} = {\hbar \over m\Omega}\delta_{mn} , ~~
\{\eta(t), \eta(t)\}_{mn} = m\Omega \hbar \delta_{mn} , 
\label{anti1} \\
&~&  \{\xi(t), \eta(t)\}_{mn} = 0 .
\label{anti2}
\end{eqnarray}
With the above, we obtain the equations of motion describing the harmonic oscillator,
\begin{eqnarray}
&~& {d \over dt} \xi_{mn}(t) = {1 \over i\hbar}[\xi, H]_{mn} = {1 \over m} \eta_{mn}(t) ,
\label{H-eq2*2x} \\
&~& {d \over dt} \eta_{mn}(t) = {1 \over i\hbar}[\eta, H]_{mn} = - m \Omega^2 \xi_{mn}(t)  ,
\label{H-eq2*2e}
\end{eqnarray}
where the Hamiltonian $H_{mn}$ is written 
\begin{eqnarray}
H_{mn} = i \Omega \sum_k \xi_{mk}(t) \eta_{kn}(t) 
 = - {1 \over 2} \hbar \Omega (\sigma^3)_{mn} .
\label{H2*2}
\end{eqnarray}

\subsection{Conjecture on M-matrix mechanics}

Let us extend the formulation described in the previous subsection to a system with M-matrix
valued quantities, whose variables are given by
\begin{eqnarray}
F_{{m_1}{m_2}\cdots{m_n}}(t) = F_{{m_1}{m_2}\cdots{m_n}} e^{i\Omega_{{m_1}{m_2}\cdots{m_n}}t} ,
\label{F}
\end{eqnarray}
where the angular frequency $\Omega_{{m_1}{m_2}\cdots{m_n}}$ 
is written in terms of antisymmetric quantities $\omega_{{m_1}{m_2}\cdots{m_{n-1}}}$ as
\begin{eqnarray}
&~& \Omega_{{m_1}{m_2}\cdots{m_n}} = \sum_{j=1}^n (-1)^{n-j}
\omega_{{m_1}\cdots{m_{j-1}}{m_{j+1}}\cdots{m_n}} \equiv
(\partial \omega)_{{m_1}{m_2}\cdots{m_n}} .
\label{Omega}
\end{eqnarray}
Here we assume the generalization of Bohrs' frequency condition\footnote{
Here and hereafter we use the reduced Planck constant $\hbar = {h \over 2\pi}$ 
as the unit of action, with the expectation that
M-matrix mechanics reduces to QM in a particular limit and 
is characterized by the same physical constant.}
\begin{eqnarray}
\Omega_{{m_1}{m_2}\cdots{m_n}} = {1 \over \hbar}\sum_{j=1}^n (-1)^{n-j}
E_{{m_1}\cdots{m_{j-1}}{m_{j+1}}\cdots{m_n}} .
\label{quanta}
\end{eqnarray}
The antisymmetric property is expressed by
\begin{eqnarray}
\Omega_{{m'_1}{m'_2}\cdots{m'_n}} = \mbox{sgn}(P) \Omega_{{m_1}{m_2}\cdots{m_n}} , ~
\omega_{{m'_1}{m'_2}\cdots{m'_{n-1}}} = \mbox{sgn}(P) \omega_{{m_1}{m_2}\cdots{m_{n-1}}} ,
\label{antisymm}
\end{eqnarray}
where sgn($P$) is $+1$ and $-1$ for even and odd permutation among indices, respectively.
The operator $\partial$ is regarded as a boundary operator 
that changes $k$-th antisymmetric objects into $(k+1)$-th objects,
and this operation is nilpotent, i.e. $\partial^2(*) =0$.\cite{hom}
Hence a homology group can be constructed from a set of phase factors of M-matrices.
The $\Omega_{{m_1}{m_2}\cdots{m_n}}$ are regarded as $(n-1)$-boundaries.
We define the hermiticity of an $n$-index object by 
$F_{{m'_1}{m'_2}\cdots{m'_n}}(t) = F_{{m_1}{m_2}\cdots{m_n}}^{*}(t)$ for odd permutations
of the subscripts.
If we define an $n$-fold product among $F_{{m_1}{m_2}\cdots{m_n}}^{(a)}(t)$
$(a = 1, 2, ...)$ by
\begin{eqnarray}
(F^{(1)} \cdots F^{(n)})_{{m_1}{m_2}\cdots{m_n}}(t) &\equiv& 
\sum_k F_{{m_1}\cdots{m_{n-1}}k}^{(1)}(t) F_{{m_1}\cdots{m_{n-2}}k{m_n}}^{(2)}(t) \cdots
F_{k{m_2}\cdots{m_{n}}}^{(n)}(t) \nonumber \\
&=& \sum_k (F^{(1)} \cdots F^{(n)})_{{m_1}{m_2}\cdots{m_n}} e^{i\Omega_{{m_1}{m_2}\cdots{m_n}}t} ,
\label{product}
\end{eqnarray}
the outcome has the same form as (\ref{F}) with the relation
$(\partial \Omega)_{{m_1}{m_2}\cdots{m_{n+1}}} = 0$,
which is a generalization of the Ritz rule.

Next, we discuss the time evolution of M-matrices, $F_{{m_1}{m_2}\cdots{m_n}}^{(a)}(t)$.
It is natural to have conjecture that the equation of motion
is given by
\begin{eqnarray}
{d \over dt}F_{{m_1}{m_2}\cdots{m_n}}^{(a)}(t) &=& i \Omega_{{m_1}{m_2}\cdots{m_n}} 
F_{{m_1}{m_2}\cdots{m_n}}^{(a)}(t) \nonumber \\
&=& {i \over \hbar}\sum_{j=1}^n (-1)^{n-j}
E_{{m_1}\cdots{m_{j-1}}{m_{j+1}}\cdots{m_n}} F_{{m_1}{m_2}\cdots{m_n}}^{(a)}(t) \nonumber \\
&=& {1 \over i\hbar} (F^{(a)}(t), K^{(1)}, \cdots, K^{(n-2)}, H)_{{m_1}{m_2}\cdots{m_n}} ,
\label{gH-eq}
\end{eqnarray}
where the quantities $K^{(1)}$, $\cdots$, $K^{(n-2)}$ and $H$ are time-independent $n$-index objects
called {\it Hamiltonians},
and $(*,*,...,*)$ is a linear combination of $n$-fold products among variables.
Equation (\ref{gH-eq}) is regarded as a generalization of the Heisenberg equation.
An ansatz for Hamiltonians and $(*,*,...,*)$ is given by
\begin{eqnarray}
&~& K^{(1)} = \cdots = K^{(n-2)} = I_{{m_1}{m_2}\cdots{m_n}} 
\nonumber \\
&~& ~~~~~~~~~~~~~ \equiv 
\sum_{(i,j)} {{{{I^{m_1 \cdots m_{i-1}}}_{m_i}}^{m_{i+1} \cdots m_{j-1}}}_{m_j}}^{m_{j+1} \cdots m_n} ,
\label{K=I} \\
&~& H_{{m_1}{m_2}\cdots{m_n}} = -{1 \over 2} \sum_{j=1}^n (-1)^{n-j} 
E_{{m_1}\cdots{m_{j-1}}{m_{j+1}}\cdots{m_n}}
\delta_{m_{j-1} m_j} 
\label{H-ansatz}
\end{eqnarray}
and
\begin{eqnarray}
&~& (F^{(1)}, F^{(2)}, \cdots, F^{(n)})_{{m_1}{m_2}\cdots{m_n}} 
= \sum_{{\scriptsize{\mbox{cyclic}}}} (F^{(1)} F^{(2)} \cdots F^{(n)})_{{m_1}{m_2}\cdots{m_n}} 
\nonumber \\
&~& ~~~~~~~~~~~~~~~~~~~~~~~~~~~~~~~~~~~~~ 
- \sum_{{\scriptsize{\mbox{cyclic}}}} (F^{(n)} \cdots F^{(2)} F^{(1)})_{{m_1}{m_2}\cdots{m_n}} ,
\label{bracket-ansatz}
\end{eqnarray}
where 
${{{{I^{m_1 \cdots m_{i-1}}}_{m_i}}^{m_{i+1} \cdots m_{j-1}}}_{m_j}}^{m_{j+1} \cdots m_n}
 = \delta_{m_i m_j} \prod_{(k,l) \neq (i,j)} (1-\delta_{m_k m_l})$ and
$\delta_{m_0 m_1} = \delta_{m_n m_1}$ in (\ref{H-ansatz}),
and  the summation in (\ref{bracket-ansatz}) is over all cyclic permutations.
The quantity $I_{{m_1}{m_2}\cdots{m_n}}$ plays the role of a unit matrix.

We now give a comment on a set of $n$-index objects.
We find that
the $(n+1) \times (n+1) \times \cdots \times (n+1)$ matrices defined by
$J^{(a)}_{{m_1}{m_2}\cdots{m_n}} \equiv -i \hbar \varepsilon_{a{m_1}{m_2}\cdots{m_n}}$
satisfy the following interesting algebra:
\begin{eqnarray}
&~& [J^{(a_1)}, J^{(a_2)}, \cdots, J^{(a_n)}]_{{m_1}{m_2}\cdots{m_n}} \nonumber \\ 
&~& ~~~~~~~~~~~~~~~~~~~~ 
= -(-1)^{n(n+1)(n-1) \over 2} (i \hbar)^{n-1} \varepsilon_{{a_1}{a_2}\cdots{a_n}{a_{n+1}}}
J^{(a_{n+1})}_{{m_1}{m_2}\cdots{m_n}} .
\label{Jan}
\end{eqnarray}
In this equation, the $n$-fold commutator is defined by
\begin{eqnarray}
&~& [F^{(a_1)}, F^{(a_2)}, \cdots, F^{(a_n)}]_{{m_1}{m_2}\cdots{m_n}} \nonumber \\
&~& ~~~~ \equiv \sum {\mbox{sgn}}(P) (F^{(a'_1)} F^{(a'_2)} \cdots F^{(a'_n)})_{{m_1}{m_2}\cdots{m_n}} ,
\label{comm-n}
\end{eqnarray}
where the summation is over all permutations among the superscripts.
The algebra (\ref{Jan}) is a generalization of ordinary spin algebra [$su(2)$ algebra]
and is equivalent to a special case of M-algebra discussed in Ref.~\cite{q2}.

\subsection{Relation to classical dynamics}

Before we study a cubic matrix, we discuss the structure of
classical dynamics from the viewpoint of matrix mechanics.
First we review the relation between CM and QM.
A physical variable $F(t)$ in CM is regarded as a linear combination of
one-index objects (a $1 \times 1$ matrix) such that
\begin{eqnarray}
F(t) = \sum_n F_n e^{i\Omega_n t} ,
\label{Fn}
\end{eqnarray}
where $F_n^* = F_{-n}$, because $F(t)$ should be a real quantity, and 
the angular frequency $\Omega_n$ is an integer multiple of the basic frequency $\omega$, 
i.e. $\Omega_n = n \omega$.
Under the guidance of Bohrs' correspondence principle and the frequency condition,
we obtain a relation between $\omega$ and the Hamiltonian $H$,
\begin{eqnarray}
\omega = {\Omega_{\Delta n} \over \Delta n}  = \lim_{{\hbar \Delta n \over n} \to 0} {\Omega_{n + \Delta n n} \over \Delta n}
= \lim_{{\hbar \Delta n \over n} \to 0} {E_{n + \Delta n} - E_{n} \over \hbar \Delta n}
= {dE \over dJ} = {\partial H \over \partial J} ,
\label{omega}
\end{eqnarray}
where $J$ is the action variable and we use $J = \oint p dq = h n$ (Bohr-Sommerfeld quantization condition).
The equation of motion for $F(t)$ is written
\begin{eqnarray}
{d \over dt}F(t) = \sum_n i n \omega F_{n} e^{i\Omega_n t} 
= {\partial F(t) \over \partial (\omega t)}{\partial H \over \partial J} 
= \{F(t), H\}_{\scriptsize{\mbox{PB}}} ,
\label{HC-eq}
\end{eqnarray}
where $\{*, *\}_{\scriptsize{\mbox{PB}}}$ is the Poisson bracket
and we use the fact that $J$ is the canonical conjugate of the angle variable $\omega t$.
Equation (\ref{HC-eq}) is Hamilton's canonical equation.

Next, we study the $\lq$classical' limit of M-matrix mechanics based on an $n$-index object, whose frequency
condition is given by (\ref{quanta}).
We require that there are generalizations of Bohr's correspondence principle and the Bohr-Sommerfeld
quantization condition and that the $\lq$classical' counterpart of the $n$-index object
satisfies Hamilton's canonical equation.
A system which satisfies these requirements is obtained under the assumption that
the variables depend on intrinsic $(n-2)$
parameters $\vec{\sigma} = (\sigma_1, ..., \sigma_{n-2})$; that is, 
a physical variable $F(t, \vec{\sigma})$ is given by
\begin{eqnarray}
F(t, \vec{\sigma}) = \sum_n F_n(\vec{\sigma}) e^{i\Omega_n t} ,
\label{Fn-sigma}
\end{eqnarray}
where $F_n^* (\vec{\sigma}) = F_{-n}(\vec{\sigma})$ and $\Omega_n = n \omega$.
The energy $E(J(\vec{\sigma}))$ is given by the functional integral
\begin{eqnarray}
E(J(\vec{\sigma})) = \int_{\Sigma} {\cal{E}}(J(\vec{\sigma})) d^{n-2} \sigma ,
\label{E-functional}
\end{eqnarray}
where $\Sigma$ is a closed $(n-2)$-dimensional surface and $J(\vec{\sigma})$ is an action variable.
The correspondence of $E(J(\vec{\sigma}))$ to $E_{{m_1}{m_2}\cdots{m_{n-1}}}$
is obtained by the replacement of $\Sigma$ with an oriented $(n-1)$-simplex:\\
$\partial({m_1}{m_2}\cdots{m_{n}})$ $= \sum_{j=1}^n (-1)^{n-j}({m_1}\cdots{m_{j-1}}{m_{j+1}}\cdots{m_{n}})$,
\begin{eqnarray} 
&~& \int_{\partial({m_1}{m_2}\cdots{m_{n}})}  {\cal{E}}(J(\vec{\sigma}))
d^{n-2} \sigma = \sum_{j=1}^n (-1)^{n-j} 
\int_{({m_1}\cdots{m_{j-1}}{m_{j+1}}\cdots{m_{n}})} {\cal{E}}(J(\vec{\sigma}))
d^{n-2} \sigma
\nonumber \\
&~& ~~~~~~~~~~~~~~~~~~~~~~~~~~~~~~~~~~ \Longleftrightarrow \sum_{j=1}^n (-1)^{n-j}
E_{{m_1}\cdots{m_{j-1}}{m_{j+1}}\cdots{m_n}} ,
\label{E-functional2}
\end{eqnarray}
that is, 
\begin{eqnarray}
\int_{({m_1}\cdots{m_{j-1}}{m_{j+1}}\cdots{m_{n}})} {\cal{E}}(J(\vec{\sigma}))
d^{n-2} \sigma \Longleftrightarrow E_{{m_1}\cdots{m_{j-1}}{m_{j+1}}\cdots{m_n}} 
\label{correspondence}
\end{eqnarray}
where $\Longleftrightarrow$ indicates the correspondence.
Generalizations of Bohr's correspondence principle and the Bohr-Sommerfeld
quantization condition are given by
\begin{eqnarray}
\omega = {\Omega_{\Delta N} \over \Delta N}  
= \lim_{\Delta N \to 0} {\Omega_{n + \Delta n_{n-1} \cdots n + \Delta n_1 n} \over \Delta N}
\label{B-Corr}
\end{eqnarray}
and 
\begin{eqnarray}
\int_{\Delta \Sigma} J(\vec{\sigma}) d^{n-2} \sigma = \hbar \Delta N ,
\label{BS-Corr}
\end{eqnarray}
respectively.
Here $\Delta N$ is a function of the quantities $\Delta n_i$ and goes to zero as the $\Delta n_i$ do,
and $\Delta \Sigma$ is an infinitesimal closed $(n-2)$-dimensional 
surface attached to a point $\vec{\sigma}$.
By use of (\ref{quanta}), (\ref{B-Corr}) and (\ref{BS-Corr}), we derive the relation
\begin{eqnarray}
&~& \omega = \lim_{\Delta N \to 0} 
{\sum_{j=1}^n (-1)^{n-j} E_{n + \Delta n_{n-1} \cdots n + \Delta n_{n-j+1} n + \Delta n_{n-j-1}
\cdots n + \Delta n_1 n} \over \hbar \Delta N} \nonumber \\
&~& ~ \Longleftrightarrow {\int_{\Delta \Sigma} {\cal{E}}(J(\vec{\sigma})) d^{n-2} \sigma
\over \int_{\Delta \Sigma} J(\vec{\sigma}) d^{n-2} \sigma} \nonumber \\
&~& ~~ = {\int_{\Delta \Sigma} ({\cal{E}}(J(\vec{\sigma}) + \Delta J(\vec{\sigma})) 
- {\cal{E}}(J(\vec{\sigma}))) d^{n-2} \sigma
\over \int_{\Delta \Sigma} \Delta J(\vec{\sigma}) d^{n-2} \sigma} 
= {\delta E(J(\vec{\sigma})) \over \delta J(\vec{\sigma})} = {\delta H \over \delta J} ,
\label{g-omega}
\end{eqnarray}
where we have used a special type of infinitesimal deformation of $J$ such that 
$\tilde{J} = J+\Delta J = 0$ on $\Delta \Sigma$ with ${\cal{E}}(\tilde{J}=0) = 0$,
and $\delta/\delta J(\vec{\sigma})$ represents the functional derivative with respect to $J(\vec{\sigma})$.
Hence, we find that $F(t, \vec{\sigma})$ satisfies Hamilton's canonical equation
$dF/dt = \{F, H\}_{\scriptsize{\mbox{PB}}}$.

Finally, we discuss the physical meaning of the $(n-2)$ parameters $\vec{\sigma}$.
The position of an object is represented by $x^i(t, \vec{\sigma})$, or
$x^{\mu}(\tau, \vec{\sigma})$ in a system with relativistic invariance.
Here, $\tau$ is a parameter that corresponds to time development and 
the $\vec{\sigma}$ are interpreted as spatial coordinates that describe an extended object.
In this way, we have arrived at the interesting conjecture that 
{\it{the $\lq$classical' counterpart of an $n$-index object is an $(n-2)$-dimensional
object and that M-matrix mechanics can describe the $\lq$quantum' physics of extended
objects.}}

\section{Cubic matrix mechanics}
 \subsection{Cubic matrix}

We now consider a three-index object (cubic matrix) given by
\begin{eqnarray}
C_{lmn}(t) = C_{lmn} e^{i\Omega_{lmn}t} ,
\label{C}
\end{eqnarray}
where the $C_{lmn}$ possesses cyclic symmetry, i.e., $C_{lmn} = C_{mnl} = C_{nlm}$, and 
the angular frequency $\Omega_{lmn}$ has the form
\begin{eqnarray}
\Omega_{lmn} = \omega_{lm} - \omega_{ln} + \omega_{mn} \equiv
(\partial \omega)_{lmn} , ~~ \omega_{ml} = -\omega_{lm} .
\label{Omegalmn}
\end{eqnarray}
The angular frequencies $\Omega_{lmn}$ have the following properties:
\begin{eqnarray}
&~& \Omega_{l'm'n'} = \mbox{sgn}(P) \Omega_{lmn} , ~~
\label{antisym}\\
&~& (\partial \Omega)_{lmnk} \equiv \Omega_{lmn} - \Omega_{lmk} + \Omega_{lnk}
 - \Omega_{mnk} = 0 .
\label{cycle}
\end{eqnarray}
The relations (\ref{Omegalmn}) and (\ref{cycle}) show that the $\Omega_{lmn}$ are 2-boundaries 

when $\partial$ is regarded as a boundary operator. 
We define the hermiticity of a cubic matrix by 
$C_{l'm'n'}(t) = C_{lmn}^{*}(t)$ for odd permutations among indices.
For a hermitian cubic matrix, there are relations
\begin{eqnarray}
C_{lmn}(t) = C_{mnl}(t) = C_{nlm}(t) = C_{mln}^{*}(t) = C_{lnm}^{*}(t) = C_{nml}^{*}(t) .
\label{hermitian}
\end{eqnarray}

If we define the triple product among cubic matrices $C_{lmn}(t) = C_{lmn} e^{i\Omega_{lmn}t}$, 
$D_{lmn}(t) = D_{lmn} e^{i\Omega_{lmn}t}$ and $E_{lmn}(t) = E_{lmn} e^{i\Omega_{lmn}t}$ by
\begin{eqnarray}
(C(t) D(t) E(t))_{lmn} &\equiv& 
\sum_k C_{lmk}(t) D_{lkn}(t) E_{kmn}(t) = (C D E)_{lmn} e^{i\Omega_{lmn}t} ,
\label{cubicproduct}
\end{eqnarray}
the product takes the same form as (\ref{C}) with the relation (\ref{cycle}).
Note that this product is, in general, neither commutative nor associative, that is,
$(CDE)_{lmn} \neq (DCE)_{lmn}$ and
$(AB(CDE))_{lmn} \neq (A(BCD)E)_{lmn} \neq ((ABC)DE)_{lmn}$.
Taking the hermitian conjugate of products for hermitian cubic matrices, 
we obtain the relations
\begin{eqnarray}
&~&(C(t) D(t) E(t))_{lmn} =  (E(t) D(t) C(t))_{nml}^{*}
 = (C(t) E(t) D(t))_{mln}^{*} \nonumber \\
&~& ~~~ = (D(t) C(t) E(t))_{lnm}^{*} = (D(t) E(t) C(t))_{nlm} = (E(t) C(t) D(t))_{mnl} .
\label{hermitianproduct}
\end{eqnarray}
The triple-commutator and anticommutator are defined by
\begin{eqnarray}
&~& [C(t), D(t), E(t)]_{lmn} \equiv (C(t)D(t)E(t) + D(t)E(t)C(t) + E(t)C(t)D(t) \nonumber \\
&~& ~~~ - D(t)C(t)E(t) - C(t)E(t)D(t) - E(t)D(t)C(t))_{lmn} 
\label{comm}
\end{eqnarray}
and
\begin{eqnarray}
&~& \{C(t), D(t), E(t)\}_{lmn} \equiv (C(t)D(t)E(t) + D(t)E(t)C(t) + E(t)C(t)D(t) \nonumber \\
&~& ~~~ + D(t)C(t)E(t) + C(t)E(t)D(t) + E(t)D(t)C(t))_{lmn} ,
\label{anticomm}
\end{eqnarray}
respectively.
With the above definitions, we have the relation
\begin{eqnarray}
&~&[A^{(a')}(t), A^{(b')}(t), A^{(c')}(t)]_{lmn} 
= \mbox{sgn}(P) [A^{(a)}(t), A^{(b)}(t), A^{(c)}(t)]_{lmn} .
\label{commR1}
\end{eqnarray}

If $C_{lmn}(t)$, $D_{lmn}(t)$ and $E_{lmn}(t)$ are hermitian matrices,
$[C(t), D(t), E(t)]_{lmn}$ and $\{C(t), D(t), E(t)\}_{lmn}$ are also hermitian cubic matrices. 

\subsection{Dynamics}

 The cyclically symmetric cubic matrices $C_{lmn}^{(a)}(t)$ yield the generalization of 
the Heisenberg equation
\begin{eqnarray}
{d \over dt}C_{lmn}^{(a)}(t) = i \Omega_{lmn} C_{lmn}^{(a)}(t) 
= {1 \over i\hbar} [C^{(a)}(t), K, H]_{lmn} ,
\label{cH-eq}
\end{eqnarray}
where $K$ and $H$ are time independent $3$-index objects.
A possible form of $K$ and $H$ is given by
\begin{eqnarray}
K_{lmn} &=& {I_{lm}}^n + {I^l}_{mn} + {{I_l}^m}_{n} \equiv I_{lmn} , 
\label{K}\\
H_{lmn} &=& {1 \over 2} \hbar \omega_{mn} {I_{lm}}^n 
+ {1 \over 2} \hbar \omega_{nl} {I^l}_{mn} + {1 \over 2} \hbar \omega_{lm} {{I_l}^m}_n  ,
\label{H} 
\end{eqnarray}
where ${I_{lm}}^n$, ${{I_l}^m}_{n}$ and ${I^l}_{mn}$ are defined by
\begin{eqnarray}
{I_{lm}}^n \equiv \delta_{lm} (1 - \delta_{nl}) , ~~
{{I_l}^m}_{n} \equiv \delta_{ln} (1 - \delta_{mn}) , ~~
{I^l}_{mn} \equiv \delta_{mn} (1 - \delta_{lm}) .
\label{Ilmn}
\end{eqnarray}

Our triple-commutator in general, does not satisfy conditions 
such as the derivation rule (which is a counterpart of the Leibniz rule in differential calculus)
and a generalization of the Jacobi identity called 
{\it a fundamental identity}, both of which are possessed by the Nambu-Poisson bracket.
As an exceptional case, the derivation rule and
the fundamental identity hold for the triple-commutator
including the Hamiltonians $K$ and $H$:
\begin{eqnarray}
{d \over dt}(C(t)D(t)E(t))_{lmn} &=& \left({dC(t) \over dt}D(t)E(t)\right)_{lmn}
+ \left(C(t){dD(t) \over dt}E(t)\right)_{lmn} \nonumber \\
&~& ~~~ + \left(C(t)D(t){dE(t) \over dt}\right)_{lmn} \nonumber \\
&=& i\Omega_{lmn}(C(t)D(t)E(t))_{lmn} \nonumber \\
&=& ([C(t), K, H]D(t)E(t))_{lmn} + (C(t)[D(t), K, H]E(t))_{lmn} \nonumber \\
&~& ~~~ + (C(t)D(t)[E(t), K, H])_{lmn}
\nonumber \\
&=& [C(t)D(t)E(t), K, H]_{lmn}  
\label{deriv}
\end{eqnarray}
for $(CDE)_{llm} = (CDE)_{lml} = (CDE)_{mll}$ and
\begin{eqnarray}
&~& [[C(t), D(t), E(t)], K, H]_{lmn} = [[C(t), K, H], D(t), E(t)]_{lmn} \nonumber \\
&~& ~~~~ + [C(t), [D(t), K, H], E(t)]_{lmn} + [C(t), D(t), [E(t), K, H]]_{lmn} .
\label{Fid}
\end{eqnarray}
It is thus seen that our description of the time development is consistent 
for cyclically symmetric matrices.

\subsection{Example}

We now study the simple example of a harmonic oscillator whose variables are two hermitian 
$3 \times 3 \times 3$ matrices
$\xi_{lmn}(t) = \xi_{lmn} e^{i\Omega_{lmn}t}$ and
$\eta_{lmn}(t) = \eta_{lmn} e^{i\Omega_{lmn}t}$. 
The coefficients $\xi_{lmn}$ and $\eta_{lmn}$ are given by
\begin{eqnarray}
\xi_{lmn} = - \sqrt{{\hbar \over 2m\Omega}} {\Omega_{lmn} \over \Omega} \varepsilon_{lmn} , ~~
\eta_{lmn} = {1 \over i}\sqrt{{m \Omega \hbar \over 2}} \varepsilon_{lmn} ,
\label{xieta3*3*3}
\end{eqnarray}
where the quantity $m$ in the square root represents a mass, and $\Omega = \Omega_{321} (>0)$.
The variables $\xi_{lmn}(t)$ and $\eta_{lmn}(t)$ satisfy the relations
\begin{eqnarray}
&~& (I \xi^2)_{lmn} = {\hbar \over 2m\Omega} {{I_{lm}}^{n}} , ~
(\xi^2 I)_{lmn} = {\hbar \over 2m\Omega} {I^l}_{mn} , ~ 
(\xi I \xi)_{lmn} = {\hbar \over 2m\Omega} {{I_l}^m}_{n} , 
\label{xi-r} \\
&~& (I \eta^2)_{lmn} = {\hbar m\Omega \over 2} {{I_{lm}}^{n}} , ~
(\eta^2 I)_{lmn} =  {\hbar m\Omega \over 2}{I^l}_{mn} , ~ 
(\eta I \eta)_{lmn} = {\hbar m\Omega \over 2} {{I_l}^m}_{n} , 
\label{eta-r} \\
&~& (I \xi \eta)_{lmn} + (I \eta \xi)_{lmn} 
= (\xi \eta I)_{lmn} + (\eta \xi I)_{lmn} 
=(\xi I \eta)_{lmn} + (\eta I \xi)_{lmn} =0 ,
\label{xieta-r1} \\
&~& (I \eta \xi)_{lmn} = {i\hbar \over 2}{I^{(3)}_{lm}}^{n} , ~ 
(\eta \xi I)_{lmn} = {i\hbar \over 2}{I^{(3)l}}_{mn} , ~
(\xi I \eta)_{lmn} = {i\hbar \over 2}{{I^{(3)}_l}^{m}}_{n} ,
\label{xieta-r2} \\
&~& (I^{(3)} \eta \xi)_{lmn} = {i\hbar \over 2}{I_{lm}}^{n} , ~ 
(\eta \xi I^{(3)})_{lmn} = {i\hbar \over 2}{I^{l}}_{mn} , ~
(\xi I^{(3)} \eta)_{lmn} = {i\hbar \over 2}{{I_l}^{m}}_{n} ,
\label{xieta-r3} \\
&~& (\xi^3)_{lmn} = (\xi^2 \eta)_{lmn} = \cdots = (\eta^2 \xi)_{lmn} = (\eta^3)_{lmn} = 0 ,
\label{r3}
\end{eqnarray}
where $I = I_{lmn}$ and 
$I^{(3)} = I^{(3)}_{lmn} \equiv {I^{(3)}_{lm}}^{n} + {I^{(3)l}}_{mn} + {{I^{(3)}_l}^{m}}_{n}$.
Here ${I^{(3)}_{lm}}^{n}$, ${I^{(3)l}}_{mn}$ and ${{I^{(3)}_l}^{m}}_{n}$ are defined by
\begin{eqnarray}
{I^{(3)}_{lm}}^{n} \equiv  \delta_{lm} \varepsilon_{mn} , ~~
{I^{(3)l}}_{mn} \equiv  \delta_{mn} \varepsilon_{nl} , ~~
{{I^{(3)}_l}^{m}}_{n} \equiv \delta_{ln} \varepsilon_{lm}  
 \label{I3} 
\end{eqnarray}
where $\varepsilon_{12} = \varepsilon_{23} = \varepsilon_{31} = - \varepsilon_{21} 
= - \varepsilon_{32} = - \varepsilon_{13} = 1$. 
With the above, we obtain the equations of motion describing the harmonic oscillator
\begin{eqnarray}
&~& {d \over dt} \xi_{lmn}(t) = {1 \over i\hbar}[\xi, K, H]_{lmn} = {1 \over m} \eta_{lmn}(t) ,
\label{H-eq3*3*3x} \\
&~& {d \over dt} \eta_{lmn}(t) = {1 \over i\hbar}[\eta, K, H]_{lmn} = - m \Omega^2 \xi_{lmn}(t) , 
\label{H-eq3*3*3e}
\end{eqnarray}
where $K$ and $H$ are given by
\begin{eqnarray}
K_{lmn} = {1 \over i \hbar}[\xi, I^{(3)}, \eta]_{lmn} = I_{lmn} , ~~
H_{lmn} = {i \over 6} \Omega [\xi, I, \eta]_{lmn} = - {1 \over 6} \hbar \Omega I^{(3)}_{lmn} .
\label{Hf} 
\end{eqnarray}

\subsection{Operator formalism}

In the preceding sections, we have studied a generalization of QM using the M-matrix formalism.
The mechanics we obtain has an interesting algebraic structure, but
the formalism is not practical, because it is only 
applicable to stationary systems. From experience,
it is known that in order to be of practical use operator formalism must be capable of
handling problems in a wider class of physical systems.
By analogy to QM, we now study the operator formalism of
cubic matrix mechanics.
First, we make the following basic assumptions.
\begin{enumerate}
\item For a given physical system, there exist triplet of state vectors 
$|m_1;{P}_{m_1m_2m_3} \rangle$, $|m_2;{P}_{m_1m_2m_3} \rangle$ and
$|m_3;{P}_{m_1m_2m_3} \rangle$
that depend on both the quantum numbers $m_i$, (e.g., these $m_i$ 
represent $l$, $m$ or $n$) and their ordering.
Here, the ordering is represented by a permutation (denoted by ${P}_{m_1m_2m_3}$) 
for a standard ordering, (e.g., $m_1 = l, m_2 = m, m_3 = n)$.

\item For every physical observable, there is a one-to-one correspondence to a linear operator $\hat{C}$.
\end{enumerate}
Under the above assumptions, it is natural to identify the cubic matrix element $C_{lmn}$
with $\hat{C} |l;{P}_{lmn} \rangle |m;{P}_{lmn} \rangle |n;{P}_{lmn} \rangle$.
In general, the quantity $C_{m_1 m_2 m_3}$ is identified with 
$\hat{C} |m_1;{P}_{m_1m_2m_3} \rangle |m_2;{P}_{m_1m_2m_3} \rangle |m_3;{P}_{m_1m_2m_3} \rangle$.
By use of (\ref{cH-eq}),
the following equations of motion for the states are derived:
\begin{eqnarray}
&~& i \hbar {d \over dt} |l;{P}_{lmn} \rangle = [\hat{K}, \hat{H}] |l;{P}_{lmn} \rangle , ~~
 i \hbar {d \over dt} |m;{P}_{lmn} \rangle = [\hat{K}, \hat{H}] |m;{P}_{lmn} \rangle , \nonumber \\
&~& i \hbar {d \over dt} |n;{P}_{lmn} \rangle = [\hat{K}, \hat{H}] |n;{P}_{lmn} \rangle .
\label{S-eq} 
\end{eqnarray}
Here, $[\hat{K}, \hat{H}]$ is the commutator of the operators $\hat{K}$ and $\hat{H}$.
(Note that $[\hat{K}, \hat{H}]$ in the third equation corresponds to 
$\sum_k (K_{lkn} H_{kmn} - H_{lkn} K_{kmn})$ in cubic matrix mechanics.)
The above equations (\ref{S-eq}) are regarded as a generalization of the Schr\"odinger equation.
The commutator ${\hat{\cal{H}}} \equiv [\hat{K}, \hat{H}]$ is interpreted as 
the generalized Hamiltonian operator.
By use of (\ref{K}) and (\ref{H}), the time evolution of state vectors is given by \begin{eqnarray}
&~& |l;{P}_{lmn} \rangle = \exp\left({i \over 2} (\omega_{nl} + \omega_{lm}) t \right)
 |l;{P}_{lmn} \rangle_0 , \nonumber \\
&~& |m;{P}_{lmn} \rangle = \exp\left({i \over 2} (\omega_{lm} + \omega_{mn}) t \right)
 |m;{P}_{lmn} \rangle_0 , \nonumber \\
&~& |n;{P}_{lmn} \rangle = \exp\left({i \over 2} (\omega_{mn} + \omega_{nl}) t \right)
 |n;{P}_{lmn} \rangle_0 ,
\label{stateslmn} 
\end{eqnarray}
where the subscript $0$ indicates that the state is that at an initial time.
In the same way, the time development of state vectors for the matrix element $C_{mln}$ is given by
\begin{eqnarray}
&~& |l;{P}_{mln} \rangle = \exp\left({i \over 2} (\omega_{ml} + \omega_{ln}) t \right)
 |l;{P}_{mln} \rangle_0 , \nonumber \\
&~& |m;{P}_{mln} \rangle = \exp\left({i \over 2} (\omega_{nm} + \omega_{ml}) t \right)
 |m;{P}_{mln} \rangle_0 , \nonumber \\
&~& |n;{P}_{mln} \rangle = \exp\left({i \over 2} (\omega_{ln} + \omega_{nm}) t \right)
 |n;{P}_{mln} \rangle_0 .
\label{statesmln} 
\end{eqnarray}
We can identify $|l;{P}_{mln} \rangle$ with the complex conjugate of $|l;{P}_{lmn} \rangle$
from (\ref{stateslmn}) and (\ref{statesmln}).
It is seen that this identification is consistent with the relations (\ref{hermitian}).

\section{Conclusions and discussion}
 We have proposed a generalization of Heisenberg's matrix mechanics
based on many-index objects.
It has been shown that there exists a solution describing a harmonic oscillator
[the three-index objects $\xi_{lmn}(t)$ and $\eta_{lmn}(t)$ defined by (\ref{xieta3*3*3})
satisfy the equations (\ref{H-eq3*3*3x}) and (\ref{H-eq3*3*3e})]
and that many-index objects lead to a generalization of spin algebra
[the $4 \times 4 \times 4$ matrices defined by
$J^{(a)}_{lmn} \equiv -i \hbar \varepsilon_{almn}$
satisfy the algebra $[J^{(a)}, J^{(b)}, J^{(c)}]_{lmn} =  \hbar^2 \varepsilon_{abcd}
J^{(d)}_{lmn}$,
where $a,b,c,d,l,m,n$ are integers from 1 to 4.]
We have studied the $\lq$classical' limit of generalized matrix mechanics and 
obtained evidence that M-matrix mechanics can be regarded as a $\lq$quantum'
theory of extended objects.
We have also made a conjecture on the operator formalism of cubic matrix mechanics.
The basic equations are given by (\ref{S-eq}).

Finally we give comments regarding the questions raised in the Introduction.

With regard to the question of why QM describes the microscopic world so successfully,
{\it the simplicity or variety of structure} in mechanics could be the key.
Quantum mechanics might represent a special case in the entirely M-matrix mechanics.
For example, matrix mechanics with many-index objects could be reduced to Heisenberg's matrix mechanics
or to a physically meaningless system by a change of variables.
It is important to make clear the entire structure of M-matrix mechanics
and find relations between its various limiting forms.

With regard to the question of whether QM holds without limit, 
there is the proposal that QM should be modified
near the Planck scale, on the basis of the problem of information loss at a black hole.\cite{BH} 
This problem is deeply related to the difficulty involved in the quantization of gravity.
Superstring theory and/or M-theory are the most promising theories
that include quantum gravity.
In fact, the problem of the counting of entropy has been solved for a class of (near) extremal black holes 
in superstring theory.\cite{BHstring}

With regard to the question of how QM is modified if it has limitations,
if elementary objects in nature
are not point particles but, rather, extended objects, the correct way to arrive at a final theory
must be to construct a theory based on a (new) mechanics appropriate for these fundamental constituents.
The study of generalized matrix mechanics might shed new light on this subject.
Or, there is the possibility that superstring theory and/or M-theory 
can be used to build a new mechanics.
It would be worthwhile to explore the generalization of QM in order to approach 
the construction of a fundamental
theory of nature from every possible direction.\footnote{
Several different ideas have been proposed for the generalization of QM.\cite{gQM}}

\end{document}